\documentclass[final,5p,times,twocolumn]{elsarticle}
\usepackage{amsmath,amssymb,bbm, ulem,enumitem,hyperref}
\usepackage{graphicx,textcomp,citesort}
\allowdisplaybreaks
\hypersetup{backref,colorlinks=true,urlcolor=blue,linkcolor=blue,citecolor=blue}

\usepackage[dvipsnames]{xcolor}

\begin{document}

\begin{frontmatter}
\title{Chiral symmetry constraints on resonant amplitudes}

\author[Regensburg]{Peter~C.~Bruns}
\author[GWU]{Maxim~Mai}

\address[Regensburg]{
Institut f\"ur Theoretische Physik, Universit\"at Regensburg,
Universit\"atsstra{\ss}e 31, 93040 Regensburg, Germany}

\address[GWU]{
Department of Physics, The George Washington University, 725 21$^{\rm st}$ St. NW, Washington, DC 20052, USA}
	\begin{abstract}
	We discuss the impact of chiral symmetry constraints on the quark-mass dependence of meson resonance pole positions, which are encoded in non-perturbative parametrizations of meson scattering amplitudes. Model-independent conditions on such parametrizations are derived, which are shown to guarantee the correct functional form of the leading quark-mass corrections to the resonance pole positions. Some model amplitudes for $\pi\pi$ scattering,  widely used for the determination of $\rho$ and $\sigma$ resonance properties from results of lattice simulations, are tested explicitly with respect to these conditions. 	
	\end{abstract}

\end{frontmatter}

\section{Introduction}\label{sec:introduction}

Chiral Perturbation Theory (ChPT) and \textit{ab-initio} lattice QCD (LQCD) simulations are currently the state of the art approaches for the exploration of low-energy QCD. The specifically beneficial overlap between these approaches arises from the fact that LQCD simulations can be (and usually are) performed at unphysical quark masses. Thus, the results of those cover the full quark mass vs. energy plane. At the same time ChPT relies on the expansion of QCD Green's functions in small momenta and quark masses, and allows for interpolations and extrapolations of the measured results in the low-energy region of that plane.

The simplest non-trivial hadronic system in this regime is the $\pi\pi$ system, which has been studied very extensively in the context of ChPT, see e.g.~\cite{ Colangelo:2001df}. Recently, high precision LQCD data became available in both $I=0$ and $I=1$ channels, see e.g.~\cite{Lang:2011mn,Dudek:2012xn, Wilson:2015dqa, Bali:2015gji, Liu:2017nzk,Guo:2016zos,Aoki:2007rd,Alexandrou:2017mpi,Briceno:2016mjc}. Usually such discrete data are extrapolated in energy and quark masses to e.g. determine properties of the (isovector) $\rho$ and (isoscalar) $\sigma$ resonances in these channels at the physical point, see~\cite{Feng:2010es, Hu:2016shf, Hu:2017wli, Doring:2016bdr, Nebreda:2010wv,Bolton:2015psa} for some recent examples. Obviously, such inter-, extrapolations require a well-founded theoretical control over the scattering amplitude in these channels. When the quark mass is fixed, the scattering amplitude is constrained by analyticity and unitarity requirements, and crossing symmetry. Many parametrizations, used in the literature, fulfill these requirements to some extent, such as the Inverse Amplitude Method, Bethe-Salpeter Equation, Breit-Wigner or Chew-Mandelstam parametrizations. The dependence on the quark masses goes beyond these requirements, and the dynamics of the underlying field theory (QCD)  has to be specified in more detail, yielding additional constraints. These constraints are mainly given by chiral symmetry, and the particular way it is broken in the real world. Close to the two-flavor chiral limit ($m_u=m_d=0$) ChPT exactly implements all these constraints, order by order in a low-energy expansion, and fixes the functional form of the quark-mass corrections to the chiral limit quantities. It is the purpose of this letter to introduce model-independent conditions on the parametrizations of the $\pi\pi$ scattering amplitude, which assure that the leading quark-mass corrections in the chiral extrapolations of the resonance properties, such as mass and width, are consistent with the chiral behavior of QCD. We find that these conditions are violated in some currently used approaches.

\section{Chiral symmetry constraints}

The effective degrees of freedom of ChPT are pseudo-Goldstone bosons (pions) of spontaneously broken chiral SU(2)$\times$SU(2) symmetry. Resonance fields can be included as explicit (massive) fields in the effective theory, see e.g. \cite{Ecker:1988te}. The corresponding Lagrangians contain bare quantities such as the bare resonance mass and couplings to pion fields, which are renormalized order-by-order in the usual sense of perturbation theory. The latter requires a proper power counting scheme, which is more subtle when massive fields are involved. The reason is that the mass and width of the meson resonances do not vanish in the chiral limit, thus introducing a new (\textit{"heavy"}) mass scale not small compared to the hadronic scale of $\sim 1\,\mathrm{GeV}$. Fortunately, one can employ tailor-made subtraction schemes which make these extended versions of the effective field theory well-defined, so that quark-mass corrections to resonance properties can be computed unambiguously, see e.g. \cite{Fuchs:2003sh,Bruns:2004tj,Djukanovic:2009zn,Bruns:2013tja,Djukanovic:2013mka}. In this letter, we focus on the purely mesonic sector of the strong interaction (in particular, $\pi\pi$ scattering), and work in the isospin-symmetric limit where $m_u=m_d=:m_{\ell}$.

Due to pion loops, any amplitude involving the strong interaction will in general depend in a non-analytic fashion on the pion mass in an expansion around the SU(2)$\times$ SU(2) chiral-symmetric limit ($m_{\ell}\to0$), as first noted in~\cite{Li:1971vr}. Denoting $M^2:=2Bm_{\ell}$, so that $M_{\pi}^{2}=M^2+\mathcal{O}(M^4\log M)$ \cite{GellMann:1968rz,Gasser:1983yg}, the mass of a \textit{"heavy"} degree of freedom in the sense explained above, denoted hereafter by $H$, depends on the light-quark mass as
\begin{equation}\label{eq:mH_expansion}
m_{H} = \overset{\circ}{m}_{H} + c_{1}^{H}M^2 + c_{2}^{H}M^3+c_{3}^{H}M^{4}\log M + \mathcal{O}(M^4)\,,
\end{equation}
where $"\circ"$ henceforth denotes the quantity (here the mass) in the chiral limit. Examples are the mass of the nucleon \cite{Gasser:1987rb}, vector meson masses \cite{Jenkins:1995vb,Bijnens:1997ni,Leinweber:2001ac,Bruns:2004tj,Djukanovic:2009zn,Bruns:2013tja} and also the mass of the kaon in a two-flavor framework where the strange-quark mass is considered as heavy compared to $m_{\ell}$~\cite{Roessl:1999iu}. An expression of the same form holds for the widths of heavy meson resonances, see e.g.~\cite{Djukanovic:2009zn,Bruns:2016zer}. We note that the non-analytic $c^H_2$-term $\sim m_{\ell}^{3/2}$ is somewhat exceptional: it exists only if there is a vertex for $H\rightarrow \pi H'$ in the effective theory, where $H'$ is mass-degenerate  with $H$ (possibly identical to $H$, or belonging to the same isospin multiplet), and is related to the threshold production of a pseudo-Goldstone boson. In the following, we will exclude this exceptional case (which is permissible in the meson sector), but we shall add some pertinent comments in the course of the investigation.
\begin{figure}[h]
\begin{center}
\includegraphics[width=0.6\linewidth]{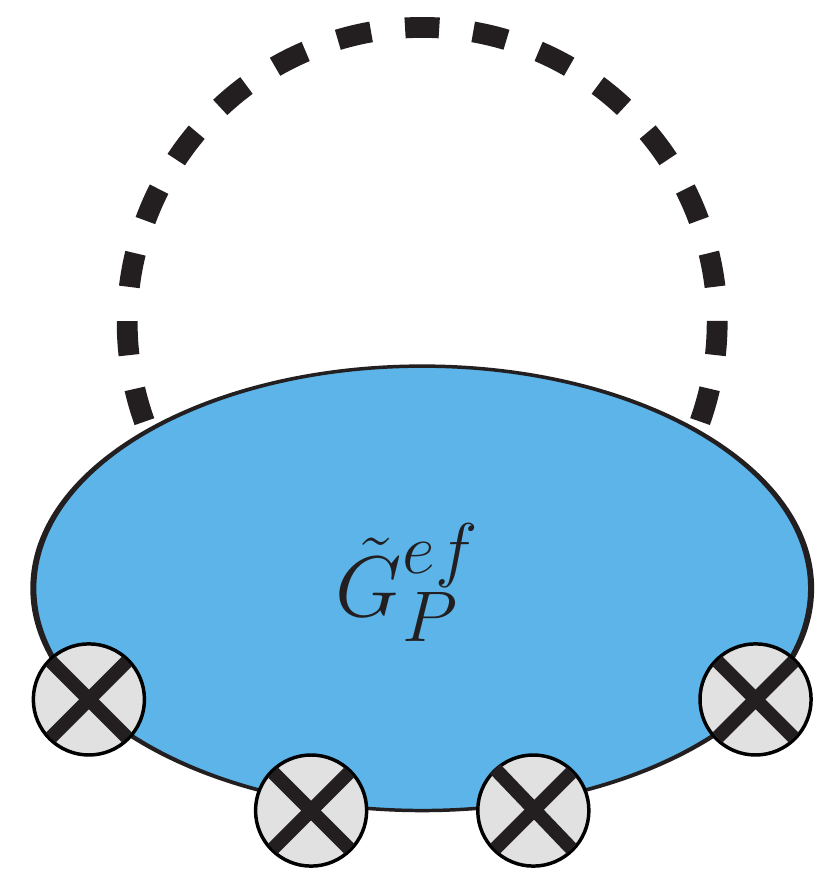}
\end{center}
\caption{Pion loop graph contributing to $G_{P}(q)$. The crossed circles denote insertions of pseudoscalar quark currents $P^{j}(x)$.}
\label{fig:GucciBag}
\end{figure}

Let us now sketch an argument showing that the quark-mass expansion of the on-shell $\pi\pi$ scattering amplitude at fixed Mandelstam variables $s,t\ne0$ contains only quark-mass logarithms with a pre\-factor of order $M^4$ or higher, to all orders in the low-energy expansion. For this purpose, we use a modified version of the general argument presented in \cite{Li:1971vr}. The $\pi^{a}(q_{a})\pi^{b}(q_{b})\to\pi^{c}(q_{c})\pi^{d}(q_{d})$ scattering amplitude can be read off from the residue of the quadruple pion pole in the Fourier transform  $G_{P}(q)$ of the correlator $\langle 0|TP^{a}(x)P^{b}(y)P^{c}(z)P^{d}(w)|0\rangle$ with respect to the space-time arguments $x,y,z,w$ \cite{Gasser:1983yg}. Here $q$ collectively denotes the pion four-mo\-men\-ta $q_{a},\ldots, q_{d}$. Moreover, let $\tilde{G}_{P}^{ef}(q,p_{e},p_{f})$ denote the analogous Fourier transform of the matrix element $\langle \pi^{f}(p_{f})|TP^{a}(x)P^{b}(y)P^{c}(z)P^{d}(w)|\pi^{e}(p_{e})\rangle$. Following the argument of \cite{Li:1971vr}, the leading quark-mass logarithm of $G_{P}(q)$ is generated by soft Goldstone bosons circulating in the loop indicated by the dashed line in Fig.~\ref{fig:GucciBag}, and can therefore be inferred from the integral
\begin{displaymath}
I_{\mathrm{log}}(q):=\frac{1}{2}\int\frac{d^{4}p}{(2\pi)^{4}}\frac{i\delta^{ef}\tilde{G}^{ef}_{P}(q,p,p)}{p^2-M^{2}}\,.
\end{displaymath}
Here $\tilde{G}_{P}^{ef}(q,p,p)$ can be taken in the ($p^{\mu}\rightarrow 0$, $M\rightarrow 0$) limit, since terms in $\tilde{G}^{ef}_{P}$ linear in $p^{\mu}$ vanish in the integral, while terms of order $p^2,M^2$ will generate terms $\sim M^4\log M^2$. In this limit, this matrix element can be expressed through four-point functions of the type of $G_{P}(q)$ in the chiral limit (and terms without a quadruple pion pole, which we can neglect here) by virtue of current algebra and PCAC techniques. Thus, the leading logarithm in the integral can be computed in terms of $\overset{\circ}{G}_{P}(q)$, employing dimensional regularization for definiteness, and scales as $\sim M^2\log M^2$. However, these logarithmic terms are exactly absorbed by the renormalization of the matrix elements $\langle 0|P^{a}|\pi^{b}\rangle = \delta^{ab}G_{\pi}$ \cite{Gasser:1983yg}, with 
\begin{displaymath}
G_{\pi}=2BF\left(1-\frac{M^2}{32\pi^2F^2}\log M^2+\ldots\right)\,,
\end{displaymath}
at the four operator insertions, and so no term ${\sim M^2\log M^2}$ is left as a correction to the remaining part of the quadruple pole term in $G_{P}(q)$, which is exactly the $\pi\pi$ scattering amplitude.
We point out that this in general requires a complicated cancellation among the Feynman graphs in the scattering amplitude, and can not be assured by power-counting arguments for individual graphs. There are some exceptions to the simple argument just given, corresponding to special cases where some combination of energy-invariants of the $\pi\pi$ process also approach zero, so that the momentum of an internal pion in $\tilde{G}^{ef}_{P}(q,p,p)$ is forced to be also ``soft'' (of order $\sim M_{\pi}$) when $p^{\mu}\rightarrow 0$. In the generic case, however, the light-quark-mass derivative of the $\pi\pi$ amplitude  exists in the chiral limit (the same is also true for the scattering of pions off heavy mesons\footnote{A consideration very similar to the one of the previous paragraph applies for matrix elements $\langle H(p')|P^{a}(x)P^{b}(y)|H(p)\rangle$, where $H$ is a heavy meson, under the qualification mentioned below Eq.~(\ref{eq:mH_expansion}). In the general case, one has to carefully analyze the Born graphs of the process $\pi\pi H \rightarrow\pi\pi H$, which is beyond the scope of this study. However, Eq.~(\ref{eq:mH_expansion}) is generally valid for heavy resonances (denoted by $R$) due to chiral symmetry and a simple power-counting argument. The considerations outlined above just serve to make plausible how this chiral-symmetry constraint is realized in resonant amplitudes $\pi\pi\rightarrow\pi\pi$ or $\pi H\rightarrow\pi H$, which do not contain the resonance degree of freedom $R$ explicitly.}). 
We have explicitly verified this constraint for the available two-loop representation for the $\pi\pi$ scattering amplitude \cite{Bijnens:1995yn} (and also for the explicit one-loop expressions for pion-kaon-scattering given in \cite{Bernard:1990kw,Bijnens:2004bu} and \cite{Roessl:1999iu}): Fixing generic non-zero energy variables $s,t,u=4M_{\pi}^{2}-s-t$ away from the $s,t$-and $u$-channel thresholds, the expansion in the light-quark mass shows only logarithmic terms with at least a pre\-factor $\sim M^4$, as a consequence of the general argument referred to above. Note that this quark-mass expansion is different from the chiral low-energy expansion, where one assumes $s,t\sim\mathcal{O}(M_{\pi}^{2})$. This is why the mentioned result is not in conflict with the corresponding one from \cite{Langacker:1973hh}, where small $s,t,u$ are presumed. We will see examples of the quark-mass expansion at fixed energy in the next section (see Eqs.~\eqref{eq:Loopexpansion}, \eqref{eq:Wexp}). 

Since the on-shell scattering amplitude shows no terms $\sim f(s,t)M^{2}\log M$ in this expansion, except for $t=0$ or $u=0$, we expect that the quark-mass expansion of the partial-wave amplitudes for $\pi\pi$ scattering will also be free of terms $\sim\tilde{f}(s)M^2\log M$, and we find that this is indeed the case. The complex-energy position $s_{H}$ of a resonance $H$ appearing in a partial wave of angular momentum $l$, $t_{l}(s)$, given by sol\-ving $(t_{l}(s_{H}))^{-1}\overset{!}{=}0$ on the second Riemann sheet in the Mandelstam variable $s$, is therefore expected to show a non-analytic quark mass dependence of $\sim M^4\log M$ or higher. This is nicely consistent with the chiral prediction of Eq.~\eqref{eq:mH_expansion} in the common case where $c^H_2=0$. Should there exist an exactly mass-degenerate resonance $H'$, with possible transitions $H\rightarrow \pi H'$ for $M_{\pi}\rightarrow 0$, the above argument must be modified, to take into account the additional $\pi H'$ branch point, and the general expectation is spoilt in this case, which makes the $c^H_2$-term necessary. But even in this case, terms $\sim M^2\log M$, which are the main concern of this study, are never present in Eq.~\eqref{eq:mH_expansion} as a consequence of chiral symmetry and chiral power-counting applied to the resonance self-energy, compare \cite{Fuchs:2003sh,Bruns:2004tj,Djukanovic:2009zn,Bruns:2013tja,Jenkins:1995vb,Bijnens:1997ni,Leinweber:2001ac,Bruns:2016zer}. The vanishing of $\tilde{f}(s)$, motivated above, will guarantee that the quark-mass dependence of the resonance position, encoded in the partial-wave amplitude, is consistent with the absence of ``forbidden logarithms'' $\sim M^2\log M$ in Eq.~(\ref{eq:mH_expansion}). This constraint can be seen as a consistency condition between two different approaches to resonances in effective field theories (explicit  inclusion of resonances, and dynamical generation of resonance poles). Thus, if a given model for the $\pi\pi$ scattering amplitude leads to such ``forbidden logarithms'', one will have to conclude that the predicted quark-mass dependence of this model is in conflict with QCD.

Following these considerations, we propose a simple test for the partial-wave amplitudes generated by a given model to fulfill the model-independent requirement, demanded by chiral symmetry -- the vanishing of terms of the form $\tilde{f}(s)M^2\log M$ in the quark-mass expansion for fixed $s\not=0$. 
Note that: 1) In the standard low-energy expansion of ChPT, this vanishing of the "forbidden logarithms" can only be verified up to a certain order $s^{n}$ in $\tilde{f}(s)$; 2) The coefficient functions, such as $\tilde{f}(s)$, in this expansion are chiral-limit quantities without quark-mass dependence, but may contain energy logarithms $\sim\log s\,$; 3) the sigma terms pertaining to the resonances, which were recently adressed as important clues to the nature of these states \cite{RuizdeElvira:2017aet}, would diverge in the chiral limit if the forbidden logarithmic terms were present in the mass formula (\ref{eq:mH_expansion}).

\section{Critical examination of model amplitudes}
\label{sec:test}

Practically all currently used model amplitudes for $\pi\pi$ scattering are of the general form  
\begin{align}\label{eq:partialwave}
t_l(s)^{-1} &= 16\pi(K_l^{-1}(s)+I(s))\,,
\end{align}
where $K_l(s)$ is a real-valued function for $0<s<16M_{\pi}^{2}$, usually referred to as $K$-matrix in cases with more channels, or generalized potential, and $I(s)$ is the two-pion loop function. 
Note that the requirement of elastic unitarity fixes
$\mathrm{Im}(16\pi I(s))=-2q(s)/\sqrt{s}$ 
for real $s>4M^{2}$, where $q(s)=\sqrt{s/4-M^2}$, such that the form of the loop function is fixed (requiring the appropriate analytic properties) up to a real constant, which can be absorbed in $K_l(s)$. In dimensional regularization with $\widetilde{MS}$ subtraction (also employed in \cite{Gasser:1983yg}) the loop function reads
\begin{align}\label{eq:I}
16\pi^2I(s)= \log\left(\frac{M^{2}}{\mu^2}\right)-1 -\frac{4q(s)}{\sqrt{s}}\mathrm{artanh}\left(\frac{-\sqrt{s}}{2q(s)}\right)\,.
\end{align}
In any channel the resonance-pole positions $s_l^*$ on the second Riemann sheet are determined as the solutions of the equation 
\begin{align}\label{eq:polcond}
{K_l^{-1}(s_l^*)+I^{II}(s_l^*) = 0}\,
\end{align} 
for ${I^{II}(s) = I(s)-iq(s)}/(4\pi\sqrt{s})$. Expanding $I^{II}(s)$ in powers of $M$ for $0\leq 4M^{2}<|s|$, one finds
\begin{align}\label{eq:Loopexpansion}
16\pi^2I^{II}(s)=
-&\left(2\pi i + 1 + \log\left(-\frac{\mu^2}{s}\right)\right)\\
+ \frac{2M^{2}}{s}&\left(2\pi i - 1 + \log\left(-\frac{M^2}{s}\right)\right)+\mathcal{O}\left(M^{4}/s^2\right)\,.  \nonumber
\end{align}
The generalized potential $K_l(s)$ parametrizes the interaction of two pions in the corresponding channel, and can be chosen in various ways. Frequently utilized examples are contact interactions from the next-to-leading chiral Lagrangian~\cite{Oller:1998hw,Hu:2016shf,Hu:2017wli}, full (including $u$ and $t$-channel loops) chiral amplitude of the next-to-leading order~\cite{Doring:2016bdr,Hanhart:2008mx,Bolton:2015psa}, or phenomenological Chew-Mandelstam forms~\cite{Dudek:2012xn,Guo:2012hv}. The most general form of the expansion of such a parametrization in powers of $M$, for fixed $s\not=0$, reads 
\begin{align}\label{eq:Wexp}
K_l^{-1}(s) = &\,\,\omega_l^{(0)}(s) 
+ \omega_l^{(1)}(s)M^{2} \\ 
&+ \omega_l^{(2)}(s)M^{2}\log\left(M^{2}/\mu^2\right)
+ \mathcal{O}(M^{4}\log M)\,. \nonumber
\end{align}
The scale dependence may partly cancel with $\omega_l^{1}(s)$ and the first term of Eq.~\eqref{eq:I}. The following condition,
\begin{equation}\label{eq:Kinvcond}
\omega_l^{(2)}(s)\overset{!}{=}-(8\pi^2s)^{-1}\,,
\end{equation}
ensures the exact cancellation of the "forbidden logarithms" in Eq.~(\ref{eq:partialwave}).

An alternative approach, not limited to models of the form of Eq.~(\ref{eq:partialwave}), is given as follows.
Insert an ansatz ${s_{H}=(m_{H}-i\Gamma_{H}/2)^2}$, with 
\begin{align}\label{eq:mHgH}
m_{H}(M) & =\overset{\circ}{m}_{H} + c_{1m}^{H}M^2 + d_{m}^{H}M^2\log M^2 +\mathcal{O}(M^3)\,,\nonumber \\
\Gamma_{H}(M) & =\overset{\circ}{\Gamma}_{H} + c_{1\Gamma}^{H}M^2 + d_{\Gamma}^{H}M^2\log M^2 +\mathcal{O}(M^3)
\end{align}
for the resonance pole position $s_l^*$ in Eq.~(\ref{eq:polcond}).
The resulting equation must hold separately in every order in $M$, since it is nothing than the resonance pole condition at any given quark mass. Now expand this resulting equation in $M$, truncate the expansion after the terms quadratic in $M$, and solve (e.g. numerically) the obtained set of equations (two for each order $M^0,M^2,M^2\log M^2$) for the unknowns $\overset{\circ}{m}_{H},\overset{\circ}{\Gamma}_{H},c_{1m}^{H},c_{1\Gamma}^{H},d_{m}^{H},d_{\Gamma}^{H}\,$. This procedure was also employed in \cite{Bruns:2016zer} for the propagator of the $\sigma$ resonance. Should the solution return non-vanishing $d_{m}^{H},d_{\Gamma}^{H}$ (the coefficients of the ``forbidden logarithms'' in $s_{H}$), the assumed model for the partial-wave amplitudes is in conflict with the strictures of chiral symmetry encoded in Eq.~(\ref{eq:mH_expansion}). In all cases examined below, we find that both versions of the test for "forbidden logarithms" are equivalent, i.e.
\begin{equation}
d_{m}^{H},d_{\Gamma}^{H} \not=0 \quad\Longleftrightarrow\quad \omega_l^{(2)}(s)\not=-(8\pi^2s)^{-1}\,.
\end{equation}
Thus, using one or another approach for the test might be a matter of technical advantages. However, when any of those fails, the model should not be used to predict the quark-mass variation of the resonance parameters, even if it describes the experimentally measured energy-dependence of the $\pi\pi$ scattering process reasonably well. This is the main point we want to make in this contribution.

As a first explicit demonstration let us adopt a ``unitarized Weinberg term'', which leads to 
\begin{align*}
K_{0}^{I=0}=\frac{2s-M_{\pi}^{2}}{2F_{\pi}^{2}}\,,~
K_{1}^{I=1}=\frac{s-4M_{\pi}^{2}}{6F_{\pi}^{2}} 
\end{align*} 
for the channels of different isospin $I$. 
Taking into account the known quark-mass dependencies of $F_{\pi}$ and $M_{\pi}$ \cite{Gasser:1983yg}, one notes that only the isoscalar s-wave described by this model fulfills the condition (\ref{eq:Kinvcond}), while the other partial waves violate this condition, even though these amplitudes are in accord with chiral symmetry on tree level. The pertaining resonance poles can therefore not be expected to vary as prescribed by Eq.~(\ref{eq:mH_expansion}) and its analogue for the width $\Gamma_{H}$.

Another typical example, frequently used for chiral extrapolations, is the Bethe-Salpeter-like approach with driving term from a local chiral potential, see e.g. \cite{Oller:1998hw,Nieves:1998hp}. As it is used for the analysis of the isovector p-wave amplitude~\cite{Hu:2016shf,Guo:2016zos,Hu:2017wli}, the expression for $K_1(s)$ in Eq.~\eqref{eq:partialwave} reads
\begin{align*}\label{eq:GWU-Kmatrix}
K_1^{\rm BSE}(s)=
\frac{32\pi q(s)^2}{48\pi(F^2-8M_\pi^2\hat l_1+4s\hat l_2)+2q(s)^2(a(\mu)+1)}
\,,
\end{align*}
where $a(\mu)$ is a real-valued subtraction constant, and $\hat l_i$ are some linear combinations of SU(3) low-energy constants (see e.g. App.~B of \cite{Guo:2016zos}). Similar to the result for $K_{1}^{I=1}(s)$ in the first model studied above, we find that Eq.~(\ref{eq:Kinvcond}) is not obeyed here (whether one takes into account the running of $F_{\pi}=F+\mathcal{O}(M^2\log M)$ with the quark mass or not), and that the $\rho$ mass and width in this model show ``forbidden logarithms''. We conclude that this kind of models can be useful when they are applied at a fixed pion mass, with their free parameters fitted at each pion mass data point separately, since the energy-dependence in the low-energy region is expected to be described reasonably well by such models. However, the quark-mass dependence of the resonance position close to the chiral limit is incompatible with the one demanded by chiral symmetry as verified by the proposed test. Therefore, ambiguities will arise when such models are used for the purpose of chiral extrapolation and the corresponding uncertainty estimates.

Finally, we consider the Inverse Amplitude Method (IAM) in the one channel version~\cite{Truong:1988zp}, see also \cite{Pelaez:2010fj}. It can be re-written in the form of~Eq.~\eqref{eq:partialwave}, such that
\begin{align*}
K_l^{\rm IAM}(s) &= \frac{(16\pi\,t_{l}^{(2)}(s))^2}{(16\pi\,t_{l}^{(2)}(s))-(16\pi\,\tilde{t}_{l}^{\,(4)})(s)}\,,
\end{align*}
where $\tilde{t}_{l}^{\,(4)}(s) := t_{l}^{(4)}(s) + 16\pi(t_{l}^{(2)}(s))^2I(s)$, and $t_l^{(n)}$ is the partial-wave scattering amplitude of the $n^{\rm th}$ chiral order. As we have already anticipated in the previous section that there are no ``forbidden logarithms'' in $t_{l}^{(4)}(s)$, it is evident that condition (\ref{eq:Kinvcond}) is fulfilled here. Thus, the quark-mass variation of the resonance position agrees with Eq.~\eqref{eq:mH_expansion} with $c_{2}^{H}=0$. In that respect,  the use of the IAM for the purpose of studying the quark-mass dependence of resonance properties in a non-perturbative framework is  preferable. Even though it disagrees with ChPT amplitudes above a certain chiral order, the fact that the IAM uses only well-behaved, complete chiral amplitudes of a fixed order as building blocks turns out as an advantage over other ``unitarization procedures''.

There are two additional remarks we wish to make. First, some resonances become stable at high pion masses. In this regime, the quark-mass variation of the resonance position could still be described satisfyingly, since the unitarity-loop effects dominate over the variation of the quark-mass logarithms there, see e.g. \cite{Guo:2011gc} for a discussion of such effects. Second, in the case of the $\rho$ resonance, there is an additional difficulty due to the $\rho\rightarrow \pi\omega$ vertex. The $\omega$ mass is very close to the $\rho$ mass (while the difference in the widths is $\mathcal{O}(M_{\pi}^{\mathrm{phys}})$), and the quark mass expansion around the $\rho$ pole in the chiral limit might have a radius of convergence smaller than the physical pion mass. If one takes the $\rho$ and $\omega$ to be mass-degenerate in the chiral limit (as is e.g. done in \cite{Djukanovic:2009zn}) to avoid this problem, one arrives at an exceptional case $c_{2}^{\rho}\not=0$ in Eq.~\eqref{eq:mH_expansion}. It is hard to see how such a behavior could be accounted for in a simple unitarized model for the $I=1$ $\pi\pi$ scattering amplitude. For the $\sigma$, however, there is no such nearly mass-degenerate state $\sigma'$ with a $\sigma\rightarrow \pi\sigma'$ vertex, so in this case the use of such models is justified, given the model in question satisfies the constraint of Eq.~\eqref{eq:Kinvcond}.\\[-1mm]

Concluding, we propose a simple test for the scattering amplitude parametrizations used for chiral extrapolations of the isovector and isoscalar $\pi\pi$ resonances. It is formulated as a model-independent condition for amplitude parametrizations of a rather general form, consistent with elastic unitarity. This condition is an implication of the chiral symmetry breaking pattern of QCD, implemented in ChPT, and ensures the correct form of the leading quark-mass correction to the resonance pole positions. 
We have tested two models frequently used for chiral extrapolations of the $\sigma$ and $\rho$ resonance poles, and found that only one is consistent with this condition.  To select those parametrizations which pass our proposed tests will clearly reduce the model-dependence afflicting the extraction of those resonance properties from results of lattice QCD simulations.

\section*{Acknowledgments}
MM thanks the German Research Foundation (DFG, MA 7156/1) for the financial support as well as M.~D\"oring and The George Washington University for hospitality and inspiring environment during this fellowship. The work of PCB was supported by the Deutsche Forschungsgemeinschaft, SFB/Transregio 55.



\begin{thebibliography}{99}
\bibitem{Colangelo:2001df} 
  G.~Colangelo, J.~Gasser and H.~Leutwyler,
  Nucl.\ Phys.\ B {\bf 603}, 125 (2001)
  [hep-ph/0103088].
  

\bibitem{Lang:2011mn}
  C.~B.~Lang, D.~Mohler, S.~Prelovsek and M.~Vidmar,
    Phys.\ Rev.\ D {\bf 84}, no. 5, 054503 (2011)
  Erratum: [Phys.\ Rev.\ D {\bf 89}, no. 5, 059903 (2014)]
    [arXiv:1105.5636 [hep-lat]].

\bibitem{Dudek:2012xn}
  J.~J.~Dudek {\it et al.} [Hadron Spectrum Collaboration],
    Phys.\ Rev.\ D {\bf 87}, no. 3, 034505 (2013)
  Erratum: [Phys.\ Rev.\ D {\bf 90}, no. 9, 099902 (2014)]
    [arXiv:1212.0830 [hep-ph]].
 


\bibitem{Wilson:2015dqa}
  D.~J.~Wilson, R.~A.~Brice\~no, J.~J.~Dudek, R.~G.~Edwards and C.~E.~Thomas,   Phys.\ Rev.\ D {\bf 92}, no. 9, 094502 (2015)
    [arXiv:1507.02599 [hep-ph]].
   

\bibitem{Bali:2015gji}
  G.~S.~Bali {\it et al.} [RQCD Collaboration],
    Phys.\ Rev.\ D {\bf 93}, no. 5, 054509 (2016)
    [arXiv:1512.08678 [hep-lat]].
  
  
\bibitem{Alexandrou:2017mpi}
  C.~Alexandrou {\it et al.},
  Phys.\ Rev.\ D {\bf 96}, no. 3, 034525 (2017)
  [arXiv:1704.05439 [hep-lat]].
  
  
  
\bibitem{Briceno:2016mjc} 
  R.~A.~Briceno, J.~J.~Dudek, R.~G.~Edwards and D.~J.~Wilson,
  Phys.\ Rev.\ Lett.\  {\bf 118}, no. 2, 022002 (2017)
  [arXiv:1607.05900 [hep-ph]].  

\bibitem{Liu:2017nzk} 
  L.~Liu {\it et al.},
   [arXiv:1701.08961 [hep-lat]
  
  

\bibitem{Guo:2016zos} 
  D.~Guo, A.~Alexandru, R.~Molina and M.~D\"oring,
    Phys.\ Rev.\ D {\bf 94}, no. 3, 034501 (2016)
    [arXiv:1605.03993 [hep-lat]].
      
  

\bibitem{Aoki:2007rd} 
  S.~Aoki {\it et al.} [CP-PACS Collaboration],
  Phys.\ Rev.\ D {\bf 76}, 094506 (2007)
    [arXiv:0708.3705 [hep-lat]].
  
  
  

\bibitem{Feng:2010es}
  X.~Feng, K.~Jansen and D.~B.~Renner,
    Phys.\ Rev.\ D {\bf 83} (2011) 094505
    [arXiv:1011.5288 [hep-lat]].
      
  
  

\bibitem{Hu:2016shf} 
  B.~Hu, R.~Molina, M.~D\"oring and A.~Alexandru,
    Phys.\ Rev.\ Lett.\  {\bf 117}, no. 12, 122001 (2016)
    [arXiv:1605.04823 [hep-lat]].
    

\bibitem{Hu:2017wli} 
  B.~Hu, R.~Molina, M.~D\"oring, M.~Mai and A.~Alexandru,
    arXiv:1704.06248 [hep-lat].
    

\bibitem{Doring:2016bdr} 
  M.~D\"oring, B.~Hu and M.~Mai,
    arXiv:1610.10070 [hep-lat].
      
  

\bibitem{Nebreda:2010wv} 
  J.~Nebreda and J.~R.~Pelaez.,
    Phys.\ Rev.\ D {\bf 81}, 054035 (2010)
    [arXiv:1001.5237 [hep-ph]].
      

\bibitem{Bolton:2015psa} 
  D.~R.~Bolton, R.~A.~Briceno and D.~J.~Wilson,
    Phys.\ Lett.\ B {\bf 757}, 50 (2016)
    [arXiv:1507.07928 [hep-ph]].
    

\bibitem{Ecker:1988te} 
  G.~Ecker, J.~Gasser, A.~Pich and E.~de Rafael,
    Nucl.\ Phys.\ B {\bf 321}, 311 (1989).

\bibitem{Fuchs:2003sh} 
  T.~Fuchs, M.~R.~Schindler, J.~Gegelia and S.~Scherer,
    Phys.\ Lett.\ B {\bf 575}, 11 (2003)
  [hep-ph/0308006].


\bibitem{Bruns:2004tj} 
  P.~C.~Bruns and U.-G.~Mei{\ss}ner,
    Eur.\ Phys.\ J.\ C {\bf 40}, 97 (2005)
  [hep-ph/0411223]. 


\bibitem{Djukanovic:2009zn} 
  D.~Djukanovic, J.~Gegelia, A.~Keller and S.~Scherer,
    Phys.\ Lett.\ B {\bf 680}, 235 (2009)
    [arXiv:0902.4347 [hep-ph]].

\bibitem{Bruns:2013tja} 
  P.~C.~Bruns, L.~Greil and A.~Sch\"afer,
    Phys.\ Rev.\ D {\bf 88}, 114503 (2013)
  [arXiv:1309.3976 [hep-ph]].

\bibitem{Djukanovic:2013mka} 
  D.~Djukanovic, E.~Epelbaum, J.~Gegelia and U.-G.~Mei{\ss}ner,
    Phys.\ Lett.\ B {\bf 730}, 115 (2014)
  [arXiv:1309.3991 [hep-ph]].

\bibitem{Li:1971vr} 
  L.~F.~Li and H.~Pagels,
    Phys.\ Rev.\ Lett.\  {\bf 26}, 1204 (1971).
  

\bibitem{GellMann:1968rz} 
  M.~Gell-Mann, R.~J.~Oakes and B.~Renner,
    Phys.\ Rev.\  {\bf 175}, 2195 (1968).
 

\bibitem{Gasser:1983yg} 
  J.~Gasser and H.~Leutwyler,
    Annals Phys.\  {\bf 158}, 142 (1984).
 
 

\bibitem{Gasser:1987rb} 
  J.~Gasser, M.~E.~Sainio and A.~Svarc,
    Nucl.\ Phys.\ B {\bf 307}, 779 (1988).     

\bibitem{Jenkins:1995vb} 
  E.~E.~Jenkins, A.~V.~Manohar and M.~B.~Wise,
    Phys.\ Rev.\ Lett.\  {\bf 75}, 2272 (1995)
    [hep-ph/9506356].
    
    

\bibitem{Bijnens:1997ni} 
  J.~Bijnens, P.~Gosdzinsky and P.~Talavera,
    Nucl.\ Phys.\ B {\bf 501}, 495 (1997)
  [hep-ph/9704212].  

\bibitem{Leinweber:2001ac} 
  D.~B.~Leinweber, A.~W.~Thomas, K.~Tsushima and S.~V.~Wright,
    Phys.\ Rev.\ D {\bf 64}, 094502 (2001)
  [hep-lat/0104013].


\bibitem{Roessl:1999iu}
  A.~Roessl,
  Nucl.\ Phys.\ B {\bf 555}, 507 (1999)
  [hep-ph/9904230].  
  
\bibitem{Bruns:2016zer} 
  P.~C.~Bruns,
    arXiv:1610.00119 [nucl-th].
  

\bibitem{Bijnens:1995yn} 
  J.~Bijnens, G.~Colangelo, G.~Ecker, J.~Gasser and M.~E.~Sainio,
    Phys.\ Lett.\ B {\bf 374}, 210 (1996)
  [hep-ph/9511397].

  

\bibitem{Bernard:1990kw} 
  V.~Bernard, N.~Kaiser and U.-G.~Mei{\ss}ner,
  Nucl.\ Phys.\ B {\bf 357}, 129 (1991).
  
\bibitem{Bijnens:2004bu} 
  J.~Bijnens, P.~Dhonte and P.~Talavera,
    JHEP {\bf 0405}, 036 (2004)
  [hep-ph/0404150].

\bibitem{Langacker:1973hh} 
  P.~Langacker and H.~Pagels,
    Phys.\ Rev.\ D {\bf 8}, 4595 (1973).

\bibitem{RuizdeElvira:2017aet} 
  J.~Ruiz de Elvira, U.-G.~Mei{\ss}ner, A.~Rusetsky and G.~Schierholz,
    arXiv:1706.09015 [hep-lat].
 

\bibitem{Oller:1998hw} 
  J.~A.~Oller, E.~Oset and J.~R.~Pelaez,
  Phys.\ Rev.\ D {\bf 59}, 074001 (1999)
  Erratum: [Phys.\ Rev.\ D {\bf 60}, 099906 (1999)]
  Erratum: [Phys.\ Rev.\ D {\bf 75}, 099903 (2007)]
  [hep-ph/9804209].

\bibitem{Hanhart:2008mx}
  C.~Hanhart, J.~R.~Pelaez and G.~Rios,
  Phys.\ Rev.\ Lett.\  {\bf 100}, 152001 (2008) arXiv:0801.2871 [hep-ph]].
    

\bibitem{Guo:2012hv}
  P.~Guo, J.~Dudek, R.~Edwards and A.~P.~Szczepaniak,
    Phys.\ Rev.\ D {\bf 88} (2013) no.1,  014501
    [arXiv:1211.0929 [hep-lat]].

\bibitem{Nieves:1998hp} 
  J.~Nieves and E.~Ruiz Arriola,
    Phys.\ Lett.\ B {\bf 455}, 30 (1999)
  [nucl-th/9807035].
      
\bibitem{Truong:1988zp}
  T.~N.~Truong,
  Phys.\ Rev.\ Lett.\  {\bf 61}, 2526 (1988).

\bibitem{Pelaez:2010fj} 
  J.~R.~Pelaez and G.~Rios,
    Phys.\ Rev.\ D {\bf 82}, 114002 (2010)
 [arXiv:1010.6008 [hep-ph]].


\bibitem{Guo:2011gc} 
  F.-K.~Guo, C.~Hanhart, F.~J.~Llanes-Estrada and U.-G.~Mei{\ss}ner,
    Phys.\ Lett.\ B {\bf 703}, 510 (2011)
  [arXiv:1105.3366 [hep-lat]].
  
  


\end{thebibliography}
\end{document}